\documentclass{egpubl}
\usepackage{eurovis2022}

\ConferencePaper        
\SpecialIssuePaper         

\usepackage[T1]{fontenc}
\usepackage{dfadobe}  

\BibtexOrBiblatex
\usepackage[backend=biber,bibstyle=EG,citestyle=alphabetic,backref=true]{biblatex} 
\electronicVersion
\PrintedOrElectronic
\ifpdf \usepackage[pdftex]{graphicx} \pdfcompresslevel=9
\else \usepackage[dvips]{graphicx} \fi

\usepackage{microtype}                
\PassOptionsToPackage{warn}{textcomp} 
\usepackage{textcomp}                
\usepackage{mathptmx}             
\usepackage{times}                     
      
\usepackage{tabu}                      
\usepackage{booktabs} 
\usepackage{multicol}
\usepackage{multirow}
\usepackage{amsmath}
\usepackage{adjustbox}
\usepackage[normalem]{ulem}
\usepackage[moderate]{savetrees}

\usepackage{wasysym}
\usepackage{tabularx}
\usepackage{url}
\usepackage{hyperref}
\usepackage{enumitem}
\usepackage{subfigure}
\PassOptionsToPackage{table,xcdraw,dvipsnames}{xcolor}
\usepackage{tikz}

\usepackage{array}
\newcolumntype{L}[1]{>{\raggedright\let\newline\\\arraybackslash\hspace{0pt}}m{#1}}
\newcolumntype{C}[1]{>{\centering\let\newline\\\arraybackslash\hspace{0pt}}m{#1}}
\newcolumntype{R}[1]{>{\raggedleft\let\newline\\\arraybackslash\hspace{0pt}}m{#1}}

\usepackage{listings}
\usepackage{xcolor}
\usepackage[T1]{fontenc}
\usepackage[labelfont=bf,textfont=it]{caption}
\usepackage{comment}
\usepackage{parskip}

\usepackage{titlesec}
\titlespacing*{\section}{0pt}{1.25ex plus 0.25ex minus 0.25ex}{0.15ex plus 0.1ex minus 0.1ex}
\titlespacing*{\subsection}{0pt}{1ex plus 0.25ex minus 0.25ex}{0.15ex plus .1ex}

\definecolor{interactioncolor}{RGB}{255,242,204}
\definecolor{sequencecolor}{RGB}{255,230,204}
\definecolor{taskcolor}{RGB}{248,206,204}
\definecolor{reasoncolor}{RGB}{225,213,231}

\definecolor{blue}{RGB}{31,119,180}
\definecolor{lightblue}{RGB}{174,199,232}
\definecolor{orange}{RGB}{255,127,14}
\definecolor{lightorange}{RGB}{255,187,120}
\definecolor{red}{RGB}{214,39,40}
\definecolor{lightred}{RGB}{247,182,210}
\definecolor{yellow}{RGB}{240,189,39}
\definecolor{lightyellow}{RGB}{255,218,102}
\definecolor{coverage}{RGB}{0,0,255}
\definecolor{specificity}{RGB}{204,0,0}

\colorlet{punct}{red!60!black}
\definecolor{background}{HTML}{FFFFFF}
\definecolor{delim}{RGB}{20,105,176}
\colorlet{numb}{magenta!60!black}

\lstdefinelanguage{json}{
    basicstyle=\normalfont\ttfamily,
    stepnumber=1,
    numbersep=8pt,
    showstringspaces=false,
    backgroundcolor=\color{background},
    literate=
     *
      {:}{{{\color{punct}{:}}}}{1}
      {,}{{{\color{punct}{,}}}}{1}
      {\{}{{{\color{delim}{\{}}}}{1}
      {\}}{{{\color{delim}{\}}}}}{1}
      {[}{{{\color{delim}{[}}}}{1}
      {]}{{{\color{delim}{]}}}}{1},
}

\usepackage[all]{nowidow}

\newif\ifnotes
\notestrue

\title{\vspace{-2ex}A Grammar-Based Approach \\for Applying Visualization Taxonomies to Interaction Logs\vspace{-2ex}}

\author[Sneha Gathani, Shayan Monadjemi, Alvitta Ottley, Leilani Battle]
{\parbox{\textwidth}{\centering Sneha Gathani$^{1}$,
        Shayan Monadjemi$^{2}$\orcid{0000-0002-9385-5969},
        Alvitta Ottley$^{2}$\orcid{0000-0002-9485-276X},
        Leilani Battle$^{3}$
        }
        \\
{\parbox{\textwidth}{\centering $^1$University of Maryland, College Park, MD\\
         $^2$ Washington University in St. Louis, St. Louis, MO\\
        $^3$ University of Washington, Seattle, WA
       }
}
}

\begin{document}

\maketitle
\begin{abstract}
Researchers collect large amounts of user interaction data with the goal of mapping user's workflows and behaviors to their high-level motivations, intuitions, and goals. 
Although the visual analytics community has proposed numerous taxonomies to facilitate this mapping process, no formal methods exist for systematically applying these existing theories to user interaction logs. 
This paper seeks to bridge the gap between visualization task taxonomies and interaction log data by making the taxonomies more actionable for interaction log analysis.
To achieve this, we leverage structural parallels between how people express themselves through interactions and language by reformulating existing theories as \textbf{regular grammars}.
We represent interactions as \textbf{terminals} within a regular grammar, similar to the role of individual words in a language, and patterns of interactions or \textbf{non-terminals} as \textbf{regular expressions} over these terminals to capture common language patterns.
To demonstrate our approach, we generate regular grammars for seven existing visualization taxonomies and develop code to apply them to three public interaction log datasets.
In analyzing these regular grammars, we find that the taxonomies at the low-level (i.e., terminals) show mixed results in expressing multiple interaction log datasets, and taxonomies at the high-level (i.e., regular expressions) have limited expressiveness, due to primarily two challenges: inconsistencies in interaction log dataset granularity and structure, and under-expressiveness of certain terminals.
Based on our findings, we suggest new research directions for the visualization community to augment existing taxonomies, develop new ones, and build better interaction log recording processes to facilitate the data-driven development of user behavior taxonomies.
\vspace{1em}


\end{abstract}  

\section{Introduction}
\label{sec:introduction}

A clear understanding of the user's visual analytic process is critical for designing and evaluating visualization systems. To this end, the visualization community has comprehensively captured their knowledge of user's visual analytic processes via multiple theoretical frameworks, typologies, and taxonomies~\cite{amar2005low, gotz2009characterizing, brehmer2013multi}. We refer to these kinds of structures as \emph{taxonomies} in this paper. In parallel, researchers are collecting more and more interaction log data to learn how humans analyze information via visualization systems in more data-driven ways~\cite{heer2008graphical,xu2020survey,cutler2020trrack,psallidas2018smoke}. The interaction log data can reveal the user's sensemaking process, analytical strategies, and reasoning behavior empirically much like taxonomies have aimed to capture them theoretically.
By enabling data-driven approaches to testing, validating, and extending longstanding theoretical taxonomies in the visualization community.

However, taxonomies generalize our understanding of user analysis behavior as \emph{high-level} user goals and strategies, whereas interaction logs aim to capture \emph{low-level} actions and system events. As a result, inferring high-level goals and analysis strategies from interaction log data often requires an explicit mapping between lower-level interactions captured with the visualization interface and a model of the user's task. One solution from the literature is to manually define user tasks based on the data and visualization system design. For example, Cook et al.~\cite{cook2015mixed} defined tasks models such as \emph{InvestigateCrime} and \emph{InvestigateSuspectsBehavior} to map low-level data interactions to potential high-level goals. This formulation enabled them to create a mixed-initiative system that infers the user's task as it evolves throughout their analysis and provides suggestions to aid the process. Similarly, Heer et al.~\cite{heer2008graphical} and Battle and Heer~\cite{battle2019characterizing} categorized their observed actions into task types such as \texttt{analysis-filter}, \texttt{undo}, \texttt{navigate}, as well as interface specific actions like \texttt{shelf-add}, \texttt{show-me} and \texttt{worksheet-add}. Customized categorizations can help researchers reveal patterns in user's analysis strategies with specific systems, but fail to generalize to other visual interfaces~\cite{psallidas2018smoke}.

Other works have leveraged existing visualization task taxonomies to systematize the analysis of collected interaction logs. For example, Pohl et al.~\cite{pohl2012analysing}, Torsney et al.~\cite{torsney2017sliceplorer} and Guo et al.~\cite{guo2015case} demonstrate the potential of using theoretical taxonomies by mapping interaction logs they gathered to pre-defined task categories. Their process first involved selecting the most appropriate theoretical taxonomy for their collected interaction data and then transforming the taxonomy into an actionable task model encoding. All three mapped their application- and task-specific actions to the same set of abstract analytic activities proposed by Yi et al.~\cite{yi2007toward} --- \texttt{select}, \texttt{explore}, \texttt{reconfigure}, \texttt{encode}, \texttt{abstract/elaborate}, \texttt{filter}, and \texttt{connect}. Similarly, Kahng et al.~\cite{kahng2020does} characterize their collected interaction logs using Gotz \& Zhou's~\cite{gotz2009characterizing} taxonomy. 
Although visualization taxonomies were not developed for the purpose of analyzing interaction log data, these mappings still enabled researchers to verify and compare log analysis methods and determine further stages of system design. Thus, existing mappings from logs to taxonomies point to an exciting opportunity to \emph{generalize} the log analysis process. However, key decision points within this mapping process are still unclear, such as which taxonomies to use or how to translate mappings for one set of system logs to another.

In this paper, we seek to bridge the gap between high-level visualization taxonomies and low-level interaction logs programmatically. In this way, we aim to extend the applicability of taxonomies by making them more actionable on interaction log datasets collected from real-world visualization systems. Our approach draws parallels between how users express themselves through interactions with analytic systems and how they express themselves via natural language by reformulating existing visualization taxonomies as \emph{regular grammars}. We represent user's recorded interactions as \emph{terminals} within a regular grammar, similar to the role of individual words in a language, and common sequences of user interactions as \emph{non-terminals}, which are defined as regular expressions over the terminals. We demonstrate the viability of our approach by generating regular grammars for seven well-known visualization taxonomies (Amar et al.~\cite{amar2005low}, Brehmer \& Munzner~\cite{brehmer2013multi}, Gotz \& Zhou~\cite{gotz2009characterizing}, Yi et al.~\cite{yi2007toward}, Guo et al.~\cite{guo2015case}, Shneiderman~\cite{shneiderman1996eyes}, and Gotz \& Wen~\cite{gotz2009behavior}) and developing code to apply them to three publicly available interaction log datasets (Battle \& Heer~\cite{battle2019characterizing}, Liu \& Heer~\cite{liu2014effects}, and Wall~\cite{wall2020detecting}). All of our code is available in our supplemental materials\footnote{\url{https://tinyurl.com/regular-grammar-taxonomies}} and provides a starting point for future works to apply the process to other taxonomies and interaction logs.

To demonstrate the utility of our grammars, we take first steps in using them to analyze the corresponding taxonomies. Specifically, we explore two new measures for analyzing the expressiveness of visualization taxonomies: \emph{coverage} and \emph{diversity}. The coverage-based measure captures the fraction of interaction log events that can be successfully mapped into a given taxonomy. The diversity-based measure examines the frequency and variety of symbols observed after mapping the taxonomy to interaction logs. Understanding such measures for various taxonomies could enable researchers to choose suitable taxonomies for their log analyses.

In analyzing the coverage and diversity of selected taxonomies,
we find that the terminal-level taxonomies (low-level) show good coverage-based results over the selected interaction log datasets but are skewed in terms of the diversity-based results, resulting in mixed overall expressiveness and the non-terminal-level taxonomies (high-level) have limited expressiveness, primarily due to two challenges: \emph{inconsistencies in log granularity and structure} and \emph{under-expressiveness of specific terminals}. We use our initial measures of expressiveness to highlight the strengths of using existing taxonomies for interaction log analysis and the limitations of their use. Furthermore, we discuss how this framework can be used to come up with new analysis measures, enrich theoretical taxonomies and build better interaction logging mechanisms to facilitate actionable and data-driven development of user behavior models.

In summary, we make the following contributions -
\begin{itemize}[itemsep=0ex, topsep=-1.5ex]
    \item We \textbf{reformulate theoretical taxonomies as regular grammars} to make them more actionable for interaction log analysis.
    \item We demonstrate our approach by \textbf{generating regular grammars for seven taxonomies} and developing code to apply them to three real-world interaction log datasets.
    \item Based on an analysis of our derived grammars, we \textbf{suggest new research directions} for augmenting existing taxonomies, developing new ones, and generalizing log recording processes to facilitate data-driven development of user behavior models.
\end{itemize}
\section{Background}
\label{sec:background}
Although taxonomies are designed to characterize the space of user interactions, they are increasingly used for log data analysis, reinforcing the importance of taxonomies in designing and evaluating visualization tools~\cite{kerracher2017constructing}. Below we summarize the literature on visualization taxonomies and their use in log data analysis.

\begin{figure}[b]
    \centering
    \vspace{-5mm}
    \includegraphics[width=\columnwidth]{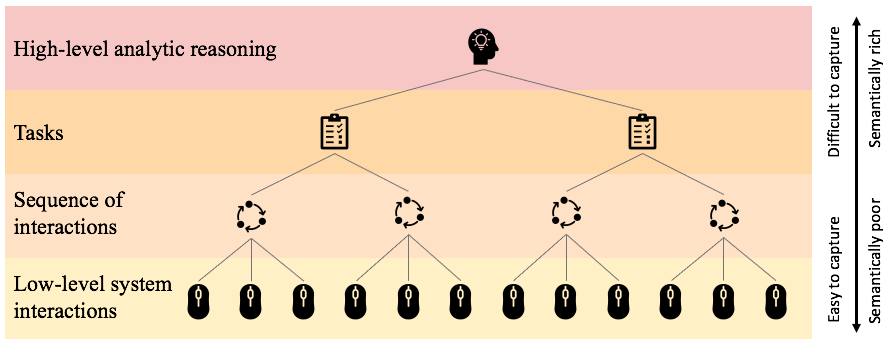}
    \caption{User interaction log data is viewed as a hierarchical construct similar to the visualization analytic activity or taxonomy structure proposed by Gotz \& Zhou~\cite{gotz2009characterizing}.}
    \label{fig:hierarchy}
\end{figure}

\subsection{Visualization Taxonomies}
Researchers generally view user interactions with visualization systems as a hierarchical construct with multiple levels of granularity, as shown in \autoref{fig:hierarchy}. These granularities are often derived in a bottom-up manner and categorized into four levels, as summarized by Gotz \& Zhou~\cite{gotz2009characterizing}: individual user interactions (e.g., \cite{sukumar2020characterizing}), sequences of user interactions (e.g., \cite{guo2015case}), user tasks (e.g., \cite{battle2019characterizing}), and high-level reasoning constructs or goals (e.g., \cite{brown2014finding,lam2017bridging}). There exist many visualization taxonomies and applications of these taxonomies at each of these four levels of granularity, which we describe below.

\textbf{Individual Interactions.} Taxonomies at the lowest-level categorize the user's most primitive interactions. Example taxonomies include those proposed by Amar et al.~\cite{amar2005low}, Yi et al.~\cite{yi2007toward} and Brehmer \& Munzner~\cite{brehmer2013multi}. Amar et al. proposed ten categories of interactions derived from explicit questions asked by students as they visually explored various datasets~\cite{amar2005low}. Yi et al. clustered the interaction capabilities of visualization tools into seven categories of interaction~\cite{yi2007toward}. Brehmer \& Munzner used a similar approach to derive a multi-level typology for user interaction~\cite{brehmer2013multi}, categorizing not only low-level interactions, but also the user's motivations for performing these interactions, such as to present information or to discover new hypotheses.

\textbf{Sequences of Interactions.} Low-level interactions are often chained together into sequences or patterns and examples of such taxonomies include ones proposed by Shneiderman~\cite{shneiderman1996eyes}, Grammel et al.~\cite{grammel2010information}, and Guo et al.~\cite{guo2015case}. Shneiderman's information-seeking mantra: ``overview first, zoom and filter, then details on demand'' describes an explicit progression of interactions performed when visually exploring data~\cite{shneiderman1996eyes}. Grammel et al. conducted a user study to understand how novices construct visualizations in Tableau~\cite{grammel2010information} and used their collected data to tabulate transitions made between attribute and encoding selection interactions. Guo et al. build on their analysis to identify sequences of interactions that frequently lead to insights in systems ~\cite{guo2015case}.

\textbf{Tasks.} Related interaction sequences can be clustered together to infer the intent of a user's analysis, often referred to as tasks. Notable examples at the task level include the taxonomies proposed by Pirolli and Card~\cite{pirolli2005sensemaking}, Battle and Heer~\cite{battle2019characterizing}, Kang et al.~\cite{kang2009evaluating}, Sedig and Parsons~\cite{sedig2013interaction}, and Alspaugh et al.~\cite{alspaugh2018futzing}. Pirolli and Card propose a pipeline of data analysis tasks encompassed within two high-level loops, foraging and sensemaking~\cite{pirolli2005sensemaking}. Battle and Heer~\cite{battle2019characterizing} survey the literature to identify common tasks completed during visual exploration. Kang et al.~\cite{kang2011characterizing}, Kandel et al.~\cite{kandel2012enterprise}, and Alspaugh et al.~\cite{alspaugh2018futzing} interview industry professionals to summarize common steps and challenges in the data analysis process. Sedig and Parsons~\cite{sedig2013interaction} use a set of tasks which are broader patterns to characterize user's mental cognitive process. Yan et al. segment sequential event logs which combines data, interaction, and user features into high-level user tasks~\cite{yan2021tessera}. Other works such as ~\cite{hibino1999task} characterize specific non-exploration tasks for information visualization of real-world data. Further, \cite{schulz2013design} characterize broad visualization tasks based on their roles such as developers, authors and end users. However, although tasks are richer semantically, they often require in-depth and arguably laborious analysis of the underlying log data to extract meaningful user activities~\cite{yan2021tessera}.

\textbf{Goals and Reasoning.} At the highest-level of the hierarchy, taxonomies aim to capture how the users \emph{organize} their analysis process into tasks and broader analysis goals. For example, Karer et al. build a formal rule-based model to reason about creation of a visualization and represent the analyst's information and knowledge flow graphically~\cite{karer2020conceptgraph}. Lam et al.~\cite{lam2017bridging} survey design study papers to understand how user's goals are broken down into actionable analysis tasks within visualization tools. Gotz et al.~\cite{gotz2006interactive} and Shrivnivasan et al.~\cite{shrinivasan2008supporting} represent analyst's mental models as links or cycles between insight discovery and knowledge understanding when tracking a user's interactions with their tools. Sedig and Parsons~\cite{sedig2013interaction} aim to speak to the user's cognitive processes via the use a patterns characterizing patterns in their analysis with visualization tools. However, we observe very few works that develop models of the user's high-level analysis intents.  We believe this stems from the challenge of modeling high-level constructs in general. Tasks and analytic reasoning structures at the highest level are difficult to capture but semantically rich as they give insight into human analytic process, while primitive interactions captured at the lowest level of the hierarchy are easier to capture but semantically poor as they provide few details about the human analytic process.

\subsection{Previous Use of Taxonomies for Interaction Log Analysis}
The literature shows that many works have used taxonomies to analyze interaction logs. For example, Guo et al. manually apply the taxonomy proposed by Yi et al.~\cite{yi2007toward} to analyze user interaction logs from a text document exploration tool~\cite{guo2015case}. Similarly, Satyanarayan et al.~\cite{satyanarayan2017vegalite} also use Yi et al.~\cite{yi2007toward} to define interaction categories to evaluate Vega-Lite. Battle and Heer~\cite{battle2019characterizing} use a similar approach to map analyst's interactions with Tableau to the Tableau-focused taxonomy of Heer et al.~\cite{heer2008graphical}. Likewise, Gotz \& Wen use the interaction taxonomy proposed by Gotz \& Zhou~\cite{gotz2009characterizing} as a means for extracting common sequences of visualization interactions~\cite{gotz2009behavior}. Battle and Scheidegger~\cite{battle2020structured} use Brehmer \& Munzner~\cite{brehmer2013multi} to guide literature review on capturing distinctions in how data management technology can be applied in interactive analysis systems. Further, Battle et al. use Pirolli and Card's sensemaking loop taxonomy~\cite{pirolli2005sensemaking} and Shneiderman's information-seeking mantra~\cite{shneiderman1996eyes} to distinguish common interaction sequences as users visually explore massive array data~\cite{battle2016dynamic}. In all cases, these applications focus less on using taxonomies for their intended use as \emph{design guidelines} and more on the unintended use of \emph{evaluation}. Only a few interaction-focused taxonomies are designed to accommodate the complex tasks observed in visual analytics~\cite{landesberger2014interaction}. Further, all these taxonomies still need to be \emph{manually} applied to individual log records, which can be a tedious process. Our goal is to work towards automation by formalizing the process of applying taxonomies to interaction logs.

\subsection {Grammar-Based Approaches to Modeling User Behavior}
Although few, there have been some interesting works in the visualization community that have taken a language-based or grammar-based perspective to understand user behavior and analytic activity. For instance, several works use Markov models~\cite{ottley2019follow, battle2016dynamic, reda2016modeling} and finite automata~\cite{dabek2016grammar} to infer user's common analysis and exploration behaviors. Dabek et al. in particular derive a grammar-based model to learn user interactions and determine common patterns for guiding new users for their visual analytic process~\cite{dabek2016grammar}. Expressing taxonomies as formal grammars can enable researchers to express theories of user analytic activity \emph{using a single, consistent language} thereby encouraging a formal means to analyze taxonomies, compare them, and either refine existing taxonomies or derive new ones. Our goal is to bridge the gap between theoretical taxonomies and data-driven analysis of visualization tools. In addition to formalizing the process of mapping taxonomies to interaction logs, we aim to enable the generation of data-driven guidelines for the design of future taxonomies \emph{and} visualization tools.
\section{Visualization Taxonomies as Regular Grammars}
\label{sec:background:taxonomies}
Inspired by ideas from linguistics and theory of computation, we leverage structural parallels between how people express themselves through interactions with analytic systems and language structure. We reformulate the hierarchical structure of visualization taxonomies (\autoref{fig:hierarchy}) as \emph{regular grammars}. Low-level user interactions can be represented as individual words in a sentence, i.e., \emph{terminal} symbols within regular grammars. More complex user behaviors captured in higher levels of the hierarchy  (sequences, tasks, goals) can be viewed as \emph{non-terminal} symbols within the regular grammar. We formulate non-terminals as \emph{regular expressions} comprised of terminals and/or other non-terminals. These regular expressions are synonymous with \emph{production rules} which are functions defined over the same terminal and/or non-terminal symbols. In this section, we motivate our approach through a concrete example of generating a regular grammar for a well-known taxonomy.

\subsection{Defining a Regular Grammar for a Taxonomy}
We define a taxonomy $t$ as being at either the low level, i.e., terminal level ($T$), or high level, i.e., non-terminal level ($NT$): $t \in \{T, NT\}$. We use $t$ to define a regular grammar, which consists of three parts:
\begin{itemize}[nosep]
    \renewcommand\labelitemi{--}
    \item a set of terminal symbols $\Sigma$,
    \item a set of non-terminal symbols $N$, and
    \item a set of production rules
    \begin{itemize}[nosep]
        \item Each production rule $f$ is a function over terminal and/or non-terminal symbols $f: N \rightarrow \{\Sigma \cup N\}^*$, where $\cup$ denotes the union of two symbol sets and $^*$ denotes repetition of items.
    \end{itemize}
\end{itemize}

The terminal symbols, $\Sigma$, are the most primitive building blocks in a grammar, i.e., the lowest level of the visualization taxonomy hierarchy in \autoref{fig:hierarchy}. Therefore, the taxonomies at the lowest level ($T$) can be seen as $\Sigma$. In the context of interaction log data, the set of distinct user interactions captured represent the primitives to be mapped to a target set of terminal symbols, similar to mapping observed words to a target dictionary.

The non-terminal symbols, $N$, capture the syntax of a grammar. These symbols resemble more complex user analysis behavior captured at higher levels of the visualization hierarchy. For example, recurring patterns or sequences of user interactions within log data can be represented as functions over terminal and non-terminal symbols, akin to deriving common sentence structures from observed word sequences. Thus taxonomies at the higher levels ($NT$) can be represented as $N$. Note that non-terminals are not limited to expressing sequences of terminals and in fact can express \emph{all levels} of the hierarchy, which we modulate through production rules. In the simplest case, non-terminals can be defined as a mapping to a single terminal, i.e., mapping a single log record to a taxonomy category. In the most complex case, non-terminals can be defined as a function of other non-terminal symbols, i.e., a function of functions. In this way, we can leverage the recursive power of regular grammars to express user analysis behavior at multiple levels of granularity.

\subsection{An Example of Generating a Regular Grammar}
\label{sec:regularlanguage:example}
\begin{figure}[t]
    \centering
    \includegraphics[width=\columnwidth]{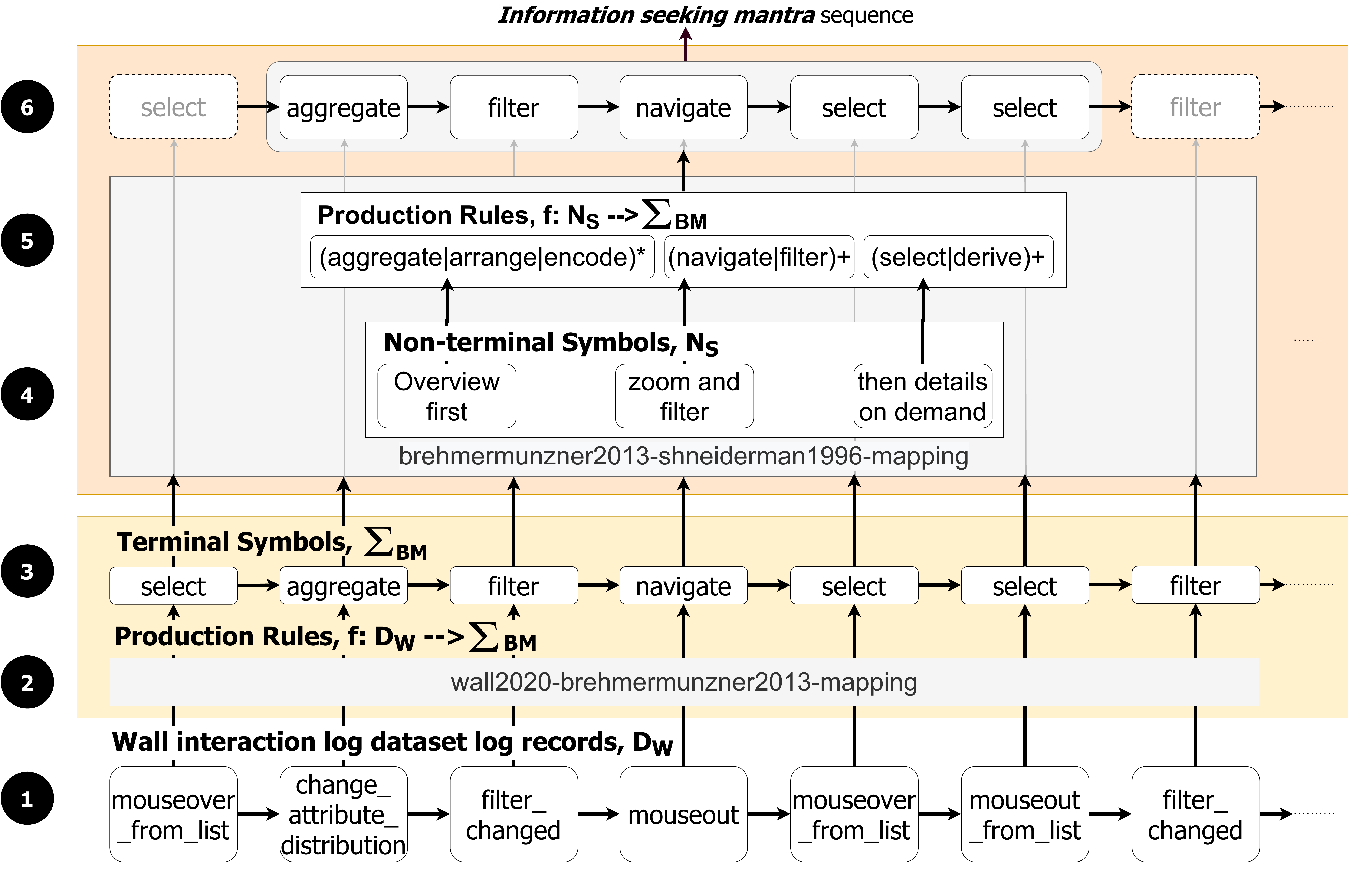}
    \caption{Example grammars for the Brehmer \& Munzner~\cite{brehmer2013multi} taxonomy and Shneiderman's information-seeking mantra~\cite{shneiderman1996eyes} applied to the Wall~\cite{wall2020detecting} interaction log dataset.}
    \label{fig:mapping_process}
    \vspace{-7mm}
\end{figure}

Here we walk through our approach of generating a regular grammar for two well-known visualization taxonomies: \emph{Brehmer \& Munzner's multi-level typology ~\cite{brehmer2013multi} ($BM$)} and \emph{Shneiderman's information-seeking mantra~\cite{shneiderman1996eyes} ($S$)}. Note that we focus on the \emph{how} level of \emph{BM}. We also demonstrate its application on an interaction log dataset collected by Wall~\cite{wall2020detecting}. This dataset captures user interactions from a visualization system intended to select a committee of politicians, enabling further study of the user's potential biases in decision making. We represent our approach in \autoref{fig:mapping_process}.

Brehmer \& Munzner classify individual interactions into 11 categories, which we represent as a set of terminal symbols $\Sigma_{BM}$:
\begin{flalign*}
 \Sigma_{BM} = \{ &\mathtt{encode}, \mathtt{select}, \mathtt{navigate}, \mathtt{arrange}, \mathtt{change}, \mathtt{filter}, && \\ 
 &\mathtt{aggregate}, \mathtt{annotate}, \mathtt{import}, \mathtt{derive}, \mathtt{record}  \} &&
\end{flalign*}
Note that we can represent any low-level visualization taxonomy as a set of terminal symbols in a similar fashion.

The Wall dataset has 11 distinct log record categories representing individual interaction types, all of which can be represented using an equivalent terminal from $\Sigma_{BM}$. As shown in \autoref{fig:mapping_process}(1), examples of these log records include \texttt{mouseover\_from\_list}, \texttt{change\_attribute\_distribution}, \texttt{filter\_changed}, etc., which allow users to retrieve details for specific politicians within the current committee list, change the rendered distribution measures computed over the politician attributes, and change the politician filtering criteria, respectively. These distinct log records can be mapped to corresponding terminals in $\Sigma_{BM}$ using production rules defined over the Wall dataset $D_W \mapsto \Sigma_{BM}$, which we label as \texttt{wall2020-brehmermunzner2013-mapping} (\autoref{fig:mapping_process}(2)). For example, we can use these functions to map the individual log records listed earlier to \texttt{select}, \texttt{aggregate}, and \texttt{filter} terminals respectively in $\Sigma_{BM}$ (\autoref{fig:mapping_process}(3)).

High-level taxonomies such as Shneiderman's information-seeking mantra (``overview first, zoom and filter, then details-on-demand''~\cite{shneiderman1996eyes}) express common sequences of user interactions observed during visual exploration. The components of this mantra (``overview'', ``zoom'', ``filter'' and ``details-on-demand'') can together be represented as a set of non-terminal symbols $N_S$ (see \autoref{fig:mapping_process}(4)):
\begin{flalign*}
 N_S = \{ &\mathtt{overview},\ \mathtt{zoom},\ \mathtt{filter},\ \mathtt{details\_on\_demand}\} &&
\end{flalign*}
Other high-level taxonomies can also be represented using non-terminal symbols in a similar fashion.

Each non-terminal in $N_S$ can be defined as a function of one or more Brehmer \& Munzner terminals: $N_S \mapsto \Sigma_{BM}$ (\autoref{fig:mapping_process}(5)), with left-hand side showing non-terminal symbol and right-hand side showing regular expression of possible terminal symbols. For example, the \texttt{overview} non-terminal symbol occurs when users transform the data, arrange the data differently or visualize the data in various ways. These transformations can be represented using the \texttt{aggregate}, \texttt{arrange} or \texttt{encode} terminals in $\Sigma_{BM}$. The corresponding production rule is as follows, defined as a regular expression:
\begin{flalign*}
\hspace{10mm}\mathtt{overview} &\rightarrow (\mathtt{aggregate | arrange | encode})^* &&
\end{flalign*}

Similarly, the \texttt{zoom}, \texttt{filter} and \texttt{details-on-demand} non-terminal symbols can be defined as regular expressions over $\Sigma_{BM}$:
\begin{flalign*}
\mathtt{zoom} &\rightarrow (\mathtt{navigate})^+ && \\
\mathtt{filter} &\rightarrow (\mathtt{filter})^+ && \\
\hspace{15mm}\mathtt{details\_on\_demand} &\rightarrow (\mathtt{select | derive})^+ && 
\end{flalign*}
We label these production rules as \texttt{brehmermunzner2013-shneiderman1996-mapping}. Similar production rules can be generated for the same non-terminal using different underlying terminals, e.g., terminals defined by Yi et al.~\cite{yi2007toward} instead of Brehmer \& Munzner~\cite{brehmer2013multi}.

Since production rules can be represented as regular expressions, we can easily apply them to user interaction sequences within logged analysis sessions to determine whether relevant patters arise within this data. For example, \autoref{fig:mapping_process}(6) shows one occurrence of the information-seeking mantra in the user session data. Similarly, other non-terminals represented by their corresponding production rules can be used to examine patterns in user interaction log data.

\section{Mapping Taxonomy Grammars to Interaction Logs}
\label{sec:approach}
\begin{table*}[t]
\centering
\caption{Statistics characterizing the three selected interaction log datasets (\emph{Battle \& Heer~\cite{battle2019characterizing}}, \emph{Liu \& Heer~\cite{liu2014effects}} and \emph{Wall~\cite{wall2020detecting}}).}
\begin{adjustbox}{width=\textwidth,center}
\begin{tabular}{c|lcccccc}
\toprule
\textbf{\begin{tabular}[c]{@{}c@{}}Interaction Log\\ Datasets\end{tabular}} & \multicolumn{1}{c}{\textbf{Datasets}} & \multicolumn{1}{l}{\textbf{\begin{tabular}[c]{@{}c@{}}Total Tasks for\\ each Participant\end{tabular}}} & \multicolumn{1}{c}{\textbf{\begin{tabular}[c]{@{}c@{}}No. of\\ Participants\end{tabular}}} & \multicolumn{1}{c}{\textbf{\begin{tabular}[c]{@{}c@{}}Min\\ Interactions\end{tabular}}} & \multicolumn{1}{c}{\textbf{\begin{tabular}[c]{@{}c@{}}Max\\ Interactions\end{tabular}}} & \multicolumn{1}{c}{\textbf{\begin{tabular}[c]{@{}c@{}}Mean of\\ Interactions\end{tabular}}} & \multicolumn{1}{c}{\textbf{\begin{tabular}[c]{@{}c@{}}Median of\\ Interactions\end{tabular}}} \\
\midrule
\rowcolor[HTML]{EFEFEF} 
\cellcolor[HTML]{EFEFEF} & flight performance & \cellcolor[HTML]{EFEFEF} & 15 & 39 & 139 & 78.6 & 84 \\
\rowcolor[HTML]{EFEFEF} 
\cellcolor[HTML]{EFEFEF} & wildlife strikes & \cellcolor[HTML]{EFEFEF} & 17 & 46 & 122 & 87.71 & 89 \\
\rowcolor[HTML]{EFEFEF} 
\multirow{-3}{*}{\cellcolor[HTML]{EFEFEF}Battle \& Heer} & weather & \multirow{-3}{*}{\cellcolor[HTML]{EFEFEF}4} & 16 & 46 & 123 & 85.19 & 84 \\
\rowcolor[HTML]{FFFFFF} 
\cellcolor[HTML]{FFFFFF} & brightkite & \cellcolor[HTML]{FFFFFF} & 16 & 8353 & 52266 & 36236.19 & 38807 \\
\rowcolor[HTML]{FFFFFF} 
\multirow{-2}{*}{\cellcolor[HTML]{FFFFFF}Liu \& Heer} & flight performance & \multirow{-2}{*}{\cellcolor[HTML]{FFFFFF}1} & 16 & 11181 & 42496 & 22969.25 & 21271.5 \\
\rowcolor[HTML]{EFEFEF} 
Wall & politicians & 1 & 24 & 122 & 783 & 370.5 & 331.5 \\
\bottomrule
\end{tabular}
\end{adjustbox}
\label{tab:stats}
\vspace{-6mm}
\end{table*}

We use best practices in qualitative coding to first map low-level interaction log records to the low-level taxonomy categories which form the terminal symbols--$\Sigma$. Then, the high-level sequences or patterns of user's behaviors established in taxonomies are represented as non-terminal symbols--$N$. Then, functions over these non-terminals are developed as production rules of combinations of simpler terminal symbols. Finally, the production rules which form regular expressions are used to observe more meaningful user sequences and patterns in the interaction log data. In order to develop our approach, we select few representative analytic visualization taxonomies and interaction log datasets which we describe first before elaborating on our approach.

\subsection{Representative Analytic Visualization Taxonomies}
We select \textbf{seven} taxonomies: \emph{four} taxonomies occurring at the lower granularity of the hierarchical structure and \emph{three} taxonomies occurring at the higher granularity as representative taxonomies on which we demonstrate the generation of regular grammar. Although we select only a subset of taxonomies here, we note that the same process can be applied directly to other taxonomies. The representative taxonomies are selected based on two factors.  First, we look for the implementation and application of the taxonomies in application systems to demonstrate their empirical use. We use the number of citations measure, with a minimum threshold of 300 as a quantitative measure of the widespreadness of the taxonomies. Second, we examine the feasibility of taxonomies such that they have enough detail to be mapped to interaction log data in order to successfully generate a regular grammar for it. We observe the number of systems that apply or use these taxonomies for either their design, analysis (e.g.,  GAN Lab~\cite{kahng2020does}) or evaluation (e.g., Sliceplorer~\cite{torsney2017sliceplorer}) as a measure of feasibility of the taxonomies. In accordance to these factors, the four representative low-level taxonomies selected are:
\begin{itemize}
    \item[] \textbf{[T1] Amar et al.'s~\cite{amar2005low}} analytic task taxonomy.
    \item[] \textbf{[T2] Brehmer \& Munzner's~\cite{brehmer2013multi}} multi-level typology of visualization tasks.
    \item[] \textbf{[T3] Gotz \& Zhou's~\cite{gotz2009characterizing}} characterization of visual activity.
    \item[] \textbf{[T4] Yi et al.'s~\cite{yi2007toward}} interaction technique category.
\end{itemize}

\noindent
And, the three representative high-level taxonomies selected are:

\begin{itemize}
    \item[] \textbf{[NT1] Shneiderman's~\cite{shneiderman1996eyes}} information-seeking mantra.
    \item[] \textbf{[NT2] Gotz \& Wen's~\cite{gotz2009behavior}} behavior patterns to infer user’s intended visual task.
    \item[] \textbf{[NT3] Guo et al.'s~\cite{guo2015case}} common analysis patterns.
\end{itemize}

\subsection{Representative Interaction Log Datasets}
We performed a search for available and pre-collected interaction log datasets online as well as reached out to our network. Following the retrieval of interaction log datasets, we specified three criteria to short-list usable datasets. First, the interaction log datasets captured needed to be of visualization-based systems. Second, the interaction log datasets needed to capture the lowest level of user interactions with visualization systems and last, the systems needed to have features that led to interaction log records mapping to at least 60\% distinct categories of the representative taxonomies. Some interaction log datasets that were ruled out due to either their unavailability or limited functionalities of the tools were Patterns and Pace dataset~\cite{feng2018patterns}, HindSight dataset~\cite{feng2016hindsight}, Anchoring Effect dataset~\cite{cho2017anchoring}, etc. We found three interaction log datasets that matched our criteria, encouraging us to apply our regular grammar's approach to them. However, our process can also be applied to other interaction log datasets as well. We list our representative datasets here, describe their tasks and provide statistics about them in \autoref{tab:stats}. 
\begin{itemize}
    \item[] \textbf{[D1] Battle \& Heer:} Data exploration using Tableau~\cite{battle2019characterizing}.
    \item[] \textbf{[D2] Liu \& Heer:} Big data exploration using imMens~\cite{liu2014effects}.
    \item[] \textbf{[D3] Wall:} Visual analytics for decision making~\cite{wall2020detecting}.
\end{itemize}

\subsection{Terminal Symbols ($\Sigma$)}
The individual user interactions in an interaction log dataset can be synonymous to low-level taxonomy categories, that is terminal symbols ($\Sigma$).

\textbf{Overview.} Therefore, for a given interaction log dataset $D$ and terminal symbol $\Sigma$, we can translate each log record $d \in D$ to its corresponding terminal symbol $i \in \Sigma$, producing the mapping $f: D \mapsto \Sigma$. For example, suppose the dragging of a slider is being mapped to the Brehmer \& Munzner terminal, represented by $\Sigma_{BM}$ (see \autoref{sec:regularlanguage:example}), $f$ may map this recorded event to the $\mathtt{filter}$ terminal $\in \Sigma_{BM}$. We develop functions over $\Sigma$ and the interaction log dataset $D$ which we call \emph{mappings}. These mappings are formulated as JSON objects, one for every interaction log dataset and low-level taxonomy that is a terminal symbol pair. Therefore, we develop a total of 12 unique mappings = 3 sets of interaction log datasets ($D$) $\times$ 4 sets of terminal taxonomies ($\Sigma_{T}$).

\textbf{Process.} 
We follow best practices in qualitative coding to derive our mappings. Similar to prior work~\cite{grammel2010information,lam2017bridging}, we used an iterative approach to establish a code-book of the mappings for all the terminal taxonomy and interaction log dataset pairs. To measure agreement among authors in applying the code-book, we calculate inter-rater reliability scores~\cite{wang2009inter}, and find that all coders were consistently in close agreement, i.e., achieved scores of 0.99 out of 1.0.

In the first iteration, for each interaction log dataset, two researchers on the project individually used the elimination approach~\cite{smith1943method} to map each distinct log record $d \in D$ to a single terminal of the low-level taxonomy, that is $i \in \Sigma$. This was repeated for all four terminal symbol taxonomies. The two researchers mapped an interaction log record to a special \emph{null} terminal if they found multiple or no terminals from the taxonomy that mapped to the log record. At the end of the first iteration, the mappings of both the researchers were aligned to calculate an inter-coder reliability score of 0.47. In order to reconcile on the code-book, the second iteration constituted of both the researchers discussing the conflicting mappings and reasoning on their choices. This discussion led to either a consensus on the mappings or finalizing the conflicts. The second iteration of forming the code-book resolved the majority of the conflicts and increased the inter-coder reliability score to 0.91. In the final round of iteration, the final conflicts were resolved by having the other two researchers on the project follow the same process which further raised the inter-coder reliability score to 0.99.

The special \emph{null} terminal was used for interaction log records that could not be mapped to a single terminal even after discussions with all four researchers. Examples of such terminals are the log records for resetting of the interface, which are commonly observed in the Tableau tool and thus in Battle \& Heer's~\cite{battle2019characterizing} interaction log dataset. The Gotz \& Zhou~\cite{gotz2009characterizing} taxonomy does not include an interaction category that resets interfaces. Therefore, we assignment \emph{null} to reset actions when mapping to the Gotz \& Zhou terminal symbol $\Sigma_{GZ}$.

We use this final iterated and established code-book of mappings (inter-coder reliability agreement = 0.99 out of 1.0) to perform further analysis. A justification for each mapping is provided as an additional nested \texttt{description} property in the JSON object mappings explaining our reasoning process. Along with the mappings, we provide a Python script that takes as input a interaction log dataset file ($D$) and the terminal symbol mapping ($\Sigma$) and outputs another file containing a list of corresponding terminal symbols, $i \in \Sigma$ of all log records of the interaction log dataset, $d \in D$.

\subsection{Non-terminal Symbols ($N$)}
The user patterns or sequences observed in interaction log datasets can be synonymous to high-level sequence taxonomy categories, that is non-terminal symbols ($N$). Therefore, we represent every high-level taxonomy describing user's sequences ($NT$) as a distinct set of non-terminal symbols, $N_{NT}$. 

\textbf{Overview.} Each pattern or sequence in a high-level taxonomy can be represented as consecutive interactions of simpler symbols, which are terminals ($\Sigma_T$). Therefore, we simply degenerate each pattern into simpler terminals and assign them to be included in the set of non-terminal symbols. These degenerations are either explicitly provided by the researchers who find these patterns or are described by them in words. Thus, based on these explicit representations or descriptions, we come up with a set of consecutive simpler terminals, $t \in NT$ that express the non-terminal sequence. Therefore, we develop a total of 3 sets of non-terminal symbols ($N_{NT}$) each consisting of one or multiple sequences or non-terminals which are represented using terminal symbols ($\Sigma_t$).

\textbf{Process.} We follow the same iterative process to establish a code-book for the non-terminal symbols mappings as followed for the terminal symbols. 
In the first iteration, two researchers on the project understand the descriptions of the founder researchers coming up with the sequences to build the set of non-terminal symbols. This is repeated for all three sets of representative non-terminal taxonomies ($N_{NT}$). A full inter-coder reliability score of 1.0 was achieved after the first iteration, thereby, saving further discussion and iterations with the other two researchers. Once again, this iterated and established code-book of non-terminals mappings was used to perform further analysis.

\subsection{Production Rules}
Empirically in interaction log datasets, complex behaviors of users are perceived to be patterns of underlying simpler interactions. In regular grammar, this can be paralleled to forming sets of non-terminal symbols ($N$) using combinations of simpler sets of terminal symbols ($\Sigma$) put together. Even more complex sequences, which speak to the more abstract tasks of the user can be formulated as combinations of not only the sets of terminal symbols but also the sets of non-terminal symbols. Our approach adopts this approach by the ability to generate production rules over non-terminal symbols. Production rules can be functions or regular expressions of both terminal and non-terminal symbols to generate complex non-terminal symbols thus informing user's common behaviors as sequences or tasks in interaction log datasets.

\textbf{Overview.} Each non-terminal can be represented as a function of terminal symbols, $\Sigma_T$. Because each low-level taxonomy or terminal symbol has a unique mapping represented by $\Sigma$, the same high-level taxonomy may be generated using different terminals, depending on which underlying low-level terminal is used. Thus, we develop functions or mappings of all possible pairings of low-level taxonomy (i.e., terminals) and high-level taxonomy (i.e., non-terminals) listed in \autoref{sec:background:taxonomies}. These mappings too are formulated as a JSON object, one for every terminal and non-terminal pair. Therefore, we develop a total of 12 unique mappings = 4 sets of terminal symbols ($\Sigma_{T}$) $\times$ 3 sets of non-terminal symbols ($N_{NT}$).

\textbf{Process.} We follow a similar iterative and best qualitative process to establish the code-book for the production rule mappings as followed for the sets of terminal and non-terminal symbols.
The mappings for non-terminal symbols are produced by building functions of mapping each individual non-terminal of the non-terminal sequence to its corresponding terminal mapping. This is repeated for all non-terminal taxonomy and terminal-taxonomy pairs. After the first iteration, where two researchers on the project individually produced the mappings for the non-terminals, a inter-coder reliability score of 0.73 was achieved. After a second iteration of discussion and resolving conflicts, a full inter-coder reliability score of 1 was achieved, thereby, saving a third iteration with the other two researchers. Once again, this final iterated and established code-book of regular expressions mappings was used to perform further analysis.

For some non-terminals, additional information was needed about the attributes and dimensions on which interactions were captured. For example, two non-terminals observed by Gotz \& Wen~\cite{gotz2009behavior}: \texttt{scan} and \texttt{drill-down} mean that users continuously perform the \texttt{inspect} iterations over a series of similar (i.e., on the same dimension or attribute) and different (i.e., on different dimensions or attributes) visual data objects respectively. To meet these needs, we modified the underlying terminal symbols to include \texttt{inspectsame} and \texttt{inspectdifferent} for \texttt{scan} and \texttt{drill-down} respectively.

Similar to the previous mappings, regular expressions also use the special \emph{null} category if the non-terminals cannot be represented as regular expressions using underlying terminals. For instance, \texttt{scan} and \texttt{drill-down} Gotz \& Wen non-terminals cannot be represented with Amar et al.~\cite{amar2005low} terminal, since there is no \texttt{inspect}-like interaction in Amar et al. taxonomy. We also provide another Python script that takes as input the terminal mappings file ($\Sigma$) and non-terminal mappings file ($N$), and outputs another file containing a list of sequences found.

\section{Analysis}
\label{sec:analysis}

\begin{table}[b]
\centering
\vspace{-5mm}
\caption{Average coverage percentage of the four sets of terminal symbols (\emph{Amar et al.}~\cite{amar2005low}, \emph{Brehmer \& Munzner}~\cite{brehmer2013multi}, \emph{Gotz \& Zhou}~\cite{gotz2009characterizing}, \emph{Yi et al.}~\cite{yi2007toward}) over three interaction log datasets (\emph{Battle \& Heer}~\cite{battle2019characterizing}, \emph{Liu \& Heer}~\cite{liu2014effects}, \emph{Wall}~\cite{wall2020detecting}).}
\begin{tabular}{c|cccc}
\toprule
\textbf{} & \textit{\textbf{Battle \& Heer}} & \textit{\textbf{Liu \& Heer}} & \textit{\textbf{Wall}} & \textbf{Avg.} \\
\midrule
\rowcolor[HTML]{EFEFEF} 
\textit{\textbf{Amar et al.}} & 46.67\% & 50\% & 100\% & 65.56\% \\
\textit{\textbf{\begin{tabular}[c]{@{}c@{}}Brehmer \&\\ Munzner\end{tabular}}} & 68.89\% & 100\% & 100\% & \textbf{89.63\%} \\
\rowcolor[HTML]{EFEFEF} 
\textit{\textbf{\begin{tabular}[c]{@{}c@{}}Gotz \&\\ Zhou\end{tabular}}} & 67.78\% & 100\% & 100\% & 89.26\% \\
\textit{\textbf{Yi et al.}} & 74.44\% & 91.67\% & 100\% & 88.7\% \\
\rowcolor[HTML]{EFEFEF} 
\textit{\textbf{Avg.}} & 64.45\% & 85.42\% & \textbf{100\%} & \\
\bottomrule
\end{tabular}
\label{tab:interactionscoverage}
\end{table}

A natural question stemming from this work is: what can we learn from translating taxonomies into regular grammars? To answer this question, we applied our derived grammars to three different interaction log datasets, and used the results to evaluate the corresponding taxonomies. Our objective is to identify \emph{measurable differences} between grammars that may reveal the \emph{most suitable taxonomies} for a given analysis context.

Inspired by prior work in evaluating visual encoding grammars~\cite{satyanarayan2017vegalite,mackinlay1986automating}, our analysis focuses on measuring the \emph{expressiveness} of selected taxonomies. We propose two measures of expressiveness for visualization taxonomies, \emph{coverage} and \emph{diversity}. When choosing taxonomies for their log analyses, we find that researchers tended to favor taxonomies where all distinct log records could be mapped to a valid terminal, i.e., taxonomies that generate mappings with high coverage of all log records. In response, our proposed \emph{coverage} measure calculates the fraction of interaction log records that can successfully be mapped to the symbols of a given taxonomy. Similarly, researchers also seemed to favor taxonomies that would avoid mapping many different log records to the same taxonomy category. For instance, if every log record maps to a single terminal, then it becomes impossible to extract meaningful interaction sequences or patterns. To this end, our \emph{diversity} measure captures the frequency and variety of symbols observed after mapping a taxonomy to the interaction log data. 

\begin{figure*}[t]
\subfigure{\includegraphics[width=\textwidth]{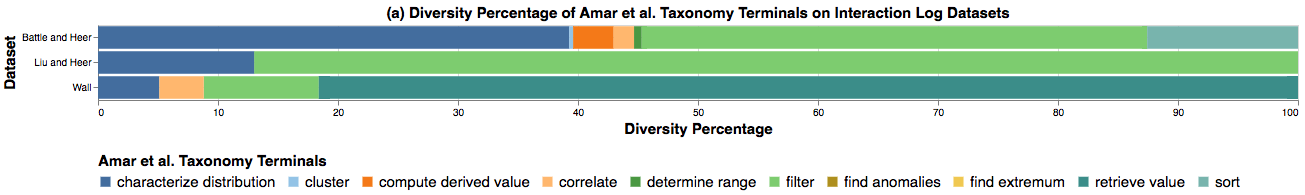}}
\vspace{-1em}
\subfigure{\includegraphics[width=\textwidth]{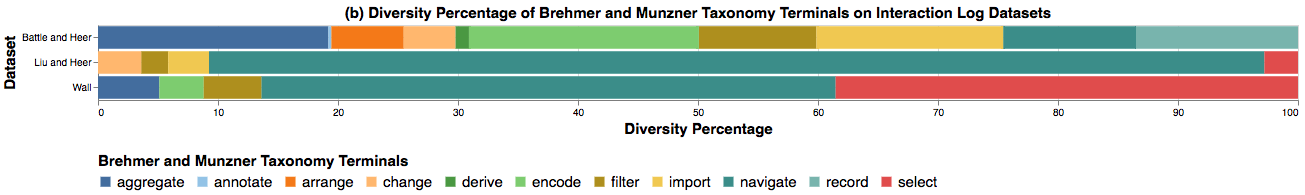}}
\vspace{-1em}
\subfigure{\includegraphics[width=\textwidth]{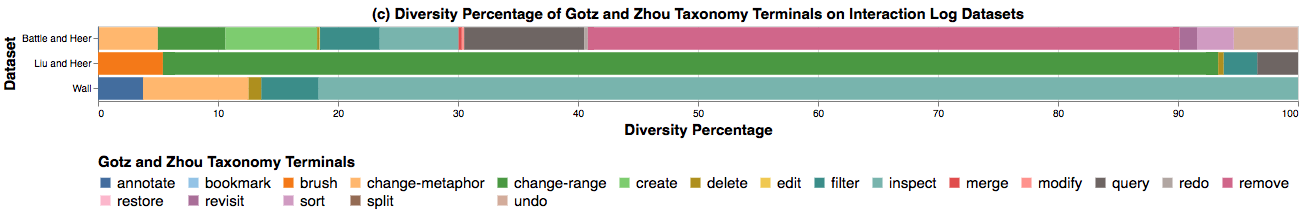}}
\vspace{-1em}
\subfigure{\includegraphics[width=\textwidth]{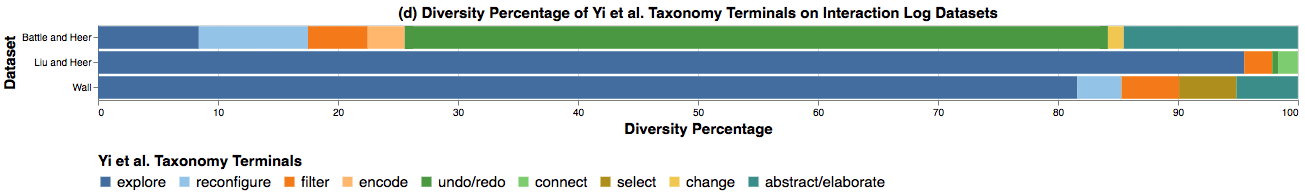}}
\vspace{.5em}
\caption{Diversity percentage for each of the terminal symbols for the four low-level taxonomies (\emph{Amar et al.}~\cite{amar2005low}, \emph{Brehmer \& Munzner}~\cite{brehmer2013multi}, \emph{Gotz \& Zhou}~\cite{gotz2009characterizing} and \emph{Yi et al.}~\cite{yi2007toward}) for the three interaction log datasets (\emph{Battle \& Heer}, \emph{Liu \& Heer}, \emph{Wall}) (top to bottom).}
\label{fig:interactionsdiversity} 
\vspace{-3mm}
\end{figure*}

\subsection{Analyzing the Expressiveness of Terminal Taxonomies}
Before we can extract patterns from log data, we first need to map it to a relevant set of terminals ($\Sigma$). However, the expressiveness of a terminal-level taxonomy can influence our ability to extract patterns, such as by having low coverage that causes us to lose data records, or by having low diversity that causes us to lose semantic meaning across interaction sequences. To evaluate the expressiveness of terminal-level taxonomies, we analyze the coverage they provide when mapped to each of our three datasets, as well as the diversity of terminals observed across the resulting mappings. We evaluate the four different representative terminal-level taxonomies in this analysis.

\subsubsection{Coverage-based Analysis}
For each terminal-level taxonomy (see \autoref{sec:background:taxonomies}), we map each distinct log record from our datasets to its corresponding terminal symbol. To measure coverage, we calculate the percentage of successfully mapped interaction log records or the percentage of ``non-null'' mappings for each interaction log dataset and taxonomy pairing, summarized in \autoref{tab:interactionscoverage}. We observe relatively high coverage for two out of three datasets: 100\% coverage of the Wall dataset, 85-100\% coverage of the Liu \& Heer dataset, and 46-68\% coverage of the Battle \& Heer dataset. We find that the Brehmer \& Munzner taxonomy~\cite{brehmer2013multi} provides the best coverage across all three datasets, followed by the taxonomy of Gotz \& Zhou~\cite{gotz2009characterizing}. The Amar et al. taxonomy~\cite{amar2005low} provided the lowest coverage, 65.56\% on average.

Number of terminals alone did not seem to be a strong predictor of coverage. For example, Brehmer \& Munzner propose fewer terminals than Gotz \& Zhou but the Brehmer \& Munzner taxonomy provides (slightly) higher coverage. Amar et al. propose fewer terminals as well, but their taxonomy provides lower coverage. Instead, we see a consistent difference in coverage based on dataset, suggesting that for the taxonomies and datasets studied, coverage is inversely proportional to the complexity of the visualization systems used to capture the interaction logs. For instance, the Wall dataset was collected using an interface with fewer features, leading to high coverage, while the Battle \& Heer dataset was captured using the feature-rich Tableau tool, resulting in the lowest observed coverage. Thus these taxonomies seem to cover simpler interfaces well but not necessarily more complex ones. We encourage further investigation in future work.

\subsubsection{Diversity-based Analysis}
To measure the diversity of our terminal-level taxonomies, we analyze the distribution of terminals observed for each interaction log dataset. Specifically, we evaluate the percentage of log records mapped to each terminal within a taxonomy to measure terminal utilization, as well as the fraction of the dataset mapped to the most popular terminal to gauge redundancies in the resulting mappings. We represent these distributions in \autoref{fig:interactionsdiversity} as four stacked bar charts, one chart per taxonomy. Each stack represents a log dataset, and each bar an individual terminal symbol from the corresponding terminal taxonomy.

\textbf{Terminal Under-Utilization.} First, we consider under-utilization of terminals, which can help us understand which taxonomies may contain terminals that do not provide meaningful context for log analyses. We find that across the four representative terminal-level taxonomies we evaluated, only two had obvious under-utilized terminals. Across the representative three log datasets we used, the Amar et al. taxonomy had two terminals that were never used: \emph{find anomalies} and \emph{find extremum}. Similarly, across all three log datasets, the Gotz \& Zhou taxonomy had three terminals that were never used: \emph{restore}, \emph{split}, and \emph{bookmark}. Further, some system actions such as \emph{formatting actions} (e.g., changing the size or color of the legends in the Tableau tool) and \emph{setting default operations} (e.g., setting default color, size, etc.) were not represented across any of the taxonomies.

\textbf{Terminal Over-Utilization.} Next, we consider over-utilization of terminals, which can help us determine when we may be losing semantic meaning across consecutive interactions. For example, if three consecutive but different interactions all mapped to the exact same terminal, we effectively lose the meaning behind these interactions---and as a consequence, the sequences derived from these interactions---in later analyses. To measure this, we identify the most popular terminal for a given log dataset and taxonomy pairing and calculate the percentage of log records mapped to this terminal. For the Battle \& Heer dataset, we find that the Brehmer \& Munzner taxonomy had the lowest measure of 19.2\% (for the \emph{aggregate} terminal), and the Gotz \& Zhou taxonomy had the highest measure of 49\% (for the \emph{remove} terminal). For the Liu \& Heer dataset, the Amar et al. taxonomy had the lowest measure of 86.99\% (for the \emph{filter} terminal), and the Yi et al. taxonomy had the highest measure of 95.43\% (for the \emph{explore} terminal). For the Wall dataset, the Brehmer \& Munzner taxonomy had the lowest measure of 47.86\% (for the \emph{navigate} terminal), whereas the Gotz \& Zhou taxonomy and Yi et al. taxonomy tied for the highest measure of 81.61\% (for the \emph{inspect} and \emph{explore} terminals, respectively). Overall, we found that the Brehmer \& Munzner taxonomy provided the best measures. We also stress that although skew in itself is not inherently bad, extreme skew, e.g., a measure of 95.43\%, means that the overwhelming majority of log records were mapped to the exact same terminal, which likely will not lead to a meaningful analysis of interaction sequences, since most interaction events will appear to be identical. If extreme skew is observed across different log datasets, then the corresponding taxonomy may not be suitable for log analysis. Further, some system actions are too broad and can override taxonomy categories. For example, the geographic filter, legend filter, filter by value, etc. in Tableau are all attributed to the \emph{filter} taxonomy category. The Wall dataset also suggests that adding or removing politicians from committee via card click or scatterplot point click are all actions that are attributed to \emph{select} taxonomy category.

We acknowledge that since none of these taxonomies were designed to analyze the specific logs we used, there are likely dataset effects that come to play in our analysis. For example, we found that terminal over-utilization was consistently high for all of our representative taxonomies when analyzing the Liu \& Heer and Wall datasets, but not for the Battle \& Heer dataset. These findings suggest that some log datasets may not be as complex as others, likely because the original interfaces themselves contained proportionally fewer features, e.g., the research prototypes developed by Liu \& Heer and Wall versus Tableau, the tool used by Battle \& Heer. That being said, we do still observe differences between taxonomies even when dataset differences are considered, with the Brehmer \& Munzner taxonomy providing overall the best over- and under-utilization results. From these skewed findings of terminals across the interaction log datasets, we see that some taxonomies have poor diversity-based measures of expressiveness. With high coverage but low diversity, the terminal taxonomies we studied show mixed overall expressiveness when applied to our selected log data.

\begin{table*}[t]
\centering
\caption{Average count of non-terminals for the three high-level taxonomies (\emph{Guo et al.}, \emph{Shneiderman} and \emph{Gotz \& Wen}) observed \textbf{per participant's session} for three interaction log datasets (\emph{Battle \& Heer}, \emph{Liu \& Heer}, \emph{Wall}) when retrieved with four underlying terminal symbols (\emph{Amar et al. (A)}, \emph{Brehmer \& Munzner (BM)}, \emph{Gotz \& Gotz (GZ)}, and \emph{Yi et al. (Y)}) with a \textbf{confidence interval of $\pm$ 95\%}.}
\begin{adjustbox}{width=\textwidth,center}
\begin{tabular}{l|l|cccccccc|cc|cccccccc}
\toprule
\multicolumn{1}{c|}{} & \multicolumn{1}{c|}{} & \multicolumn{8}{c|}{\textbf{Guo et al.}} & \multicolumn{2}{c|}{\textbf{Shneiderman}} & \multicolumn{8}{c}{\textbf{Gotz \& Wen}} \\ \cline{3-20} 
\multicolumn{1}{c|}{} & \multicolumn{1}{c|}{} & \multicolumn{2}{c}{\textit{elaborating}} & \multicolumn{2}{c}{\textit{locating}} & \multicolumn{2}{c}{\textit{orienting}} & \multicolumn{2}{c|}{\textit{sampling}} & \multicolumn{2}{c|}{\textit{ISM}} & \multicolumn{2}{c}{\textit{drill-down}} & \multicolumn{2}{c}{\textit{swap}} & \multicolumn{2}{c}{\textit{flip}} & \multicolumn{2}{c}{\textit{scan}} \\ \cline{3-20} 
\multicolumn{1}{c|}{\multirow{-3}{*}{\textbf{\begin{tabular}[c]{@{}c@{}}Provenance\\ Datasets\end{tabular}}}} & \multicolumn{1}{c|}{\multirow{-3}{*}{\textbf{\begin{tabular}[c]{@{}c@{}}Interaction\\ Taxonomy\end{tabular}}}} & \multicolumn{1}{l}{Count} & \multicolumn{1}{l}{\cellcolor[HTML]{DAE8FC} $\pm 95\%$} & \multicolumn{1}{l}{Count} & \multicolumn{1}{l}{\cellcolor[HTML]{DAE8FC} $\pm 95\%$} & \multicolumn{1}{l}{Count} & \multicolumn{1}{l}{\cellcolor[HTML]{DAE8FC} $\pm 95\%$} & \multicolumn{1}{l}{Count} & \multicolumn{1}{l|}{\cellcolor[HTML]{DAE8FC} $\pm 95\%$} & \multicolumn{1}{l}{Count} & \multicolumn{1}{l|}{\cellcolor[HTML]{DAE8FC} $\pm 95\%$} & \multicolumn{1}{l}{Count} & \multicolumn{1}{l}{\cellcolor[HTML]{DAE8FC} $\pm 95\%$} & \multicolumn{1}{l}{Count} & \multicolumn{1}{l}{\cellcolor[HTML]{DAE8FC} $\pm 95\%$} & \multicolumn{1}{l}{Count} & \multicolumn{1}{l}{\cellcolor[HTML]{DAE8FC} $\pm 95\%$} & \multicolumn{1}{l}{Count} & \multicolumn{1}{l}{\cellcolor[HTML]{DAE8FC} $\pm 95\%$} \\
\midrule
\rowcolor[HTML]{EFEFEF} 
\cellcolor[HTML]{EFEFEF} & A & 0 & \cellcolor[HTML]{E4EEFC}0 & 0 & \cellcolor[HTML]{E4EEFC}0 & 0 & \cellcolor[HTML]{E4EEFC}0 & 0 & \cellcolor[HTML]{E4EEFC}0 & 0.04 & \cellcolor[HTML]{E4EEFC}0.58 & 0 & \cellcolor[HTML]{E4EEFC}0 & 1.12 & \cellcolor[HTML]{E4EEFC}0.41 & 3 & \cellcolor[HTML]{E4EEFC}0.62 & 0 & \cellcolor[HTML]{E4EEFC}0 \\
\rowcolor[HTML]{EFEFEF} 
\cellcolor[HTML]{EFEFEF} & BM & 0 & \cellcolor[HTML]{E4EEFC}0 & 0 & \cellcolor[HTML]{E4EEFC}0 & 0 & \cellcolor[HTML]{E4EEFC}0 & 0 & \cellcolor[HTML]{E4EEFC}0 & 0 & \cellcolor[HTML]{E4EEFC}0 & 5.75 & \cellcolor[HTML]{E4EEFC}0.65 & 1.27 & \cellcolor[HTML]{E4EEFC}0.58 & 5.88 & \cellcolor[HTML]{E4EEFC}0.87 & 5.19 & \cellcolor[HTML]{E4EEFC}0.94 \\
\rowcolor[HTML]{EFEFEF} 
\cellcolor[HTML]{EFEFEF} & GZ & 2.25 & \cellcolor[HTML]{E4EEFC}0.55 & 0.62 & \cellcolor[HTML]{E4EEFC}0.28 & 0 & \cellcolor[HTML]{E4EEFC}0 & 0.17 & \cellcolor[HTML]{E4EEFC}0.11 & 0.48 & \cellcolor[HTML]{E4EEFC}0.2 & 5.19 & \cellcolor[HTML]{E4EEFC}0.89 & 1.38 & \cellcolor[HTML]{E4EEFC}0.5 & 8.23 & \cellcolor[HTML]{E4EEFC}1.22 & 4.38 & \cellcolor[HTML]{E4EEFC}0.76 \\
\rowcolor[HTML]{EFEFEF} 
\cellcolor[HTML]{EFEFEF}Battle \& Heer & Y & 6.52 & \cellcolor[HTML]{E4EEFC}1.28 & 2.62 & \cellcolor[HTML]{E4EEFC}0.59 & 0 & \cellcolor[HTML]{E4EEFC}0 & 0.38 & \cellcolor[HTML]{E4EEFC}0.21 & 0 & \cellcolor[HTML]{E4EEFC}0 & 6.33 & \cellcolor[HTML]{E4EEFC}0.83 & 4.71 & \cellcolor[HTML]{E4EEFC}1.06 & 0 & \cellcolor[HTML]{E4EEFC}0 & 5.46 & \cellcolor[HTML]{E4EEFC}1.02 \\
\rowcolor[HTML]{FFFFFF} 
\cellcolor[HTML]{FFFFFF} & A & 0 & \cellcolor[HTML]{F2F7FC}0 & 0 & \cellcolor[HTML]{F2F7FC}0 & 0 & \cellcolor[HTML]{F2F7FC}0 & 0 & \cellcolor[HTML]{F2F7FC}0 & 0 & \cellcolor[HTML]{F2F7FC}0 & 0 & \cellcolor[HTML]{F2F7FC}0 & 0 & \cellcolor[HTML]{F2F7FC}0 & 1 & \cellcolor[HTML]{F2F7FC}0 & 0 & \cellcolor[HTML]{F2F7FC}0 \\
\rowcolor[HTML]{FFFFFF} 
\cellcolor[HTML]{FFFFFF} & BM & 165.5 & \cellcolor[HTML]{F2F7FC}214.18 & 0 & \cellcolor[HTML]{F2F7FC}0 & 0 & \cellcolor[HTML]{F2F7FC}0 & 0 & \cellcolor[HTML]{F2F7FC}0 & 40.94 & \cellcolor[HTML]{F2F7FC}18.41 & 162.62 & \cellcolor[HTML]{F2F7FC}203.79 & 0 & \cellcolor[HTML]{F2F7FC}0 & 83.12 & \cellcolor[HTML]{F2F7FC}23.07 & 44 & \cellcolor[HTML]{F2F7FC}11.42 \\
\rowcolor[HTML]{FFFFFF} 
\cellcolor[HTML]{FFFFFF} & GZ & 252.19 & \cellcolor[HTML]{F2F7FC}230.16 & 22.75 & \cellcolor[HTML]{F2F7FC}5.57 & 0 & \cellcolor[HTML]{F2F7FC}0 & 5.06 & \cellcolor[HTML]{F2F7FC}2.18 & 87.38 & \cellcolor[HTML]{F2F7FC}36.75 & 152.56 & \cellcolor[HTML]{F2F7FC}53.19 & 0 & \cellcolor[HTML]{F2F7FC}0 & 141.06 & \cellcolor[HTML]{F2F7FC}48.81 & 124.88 & \cellcolor[HTML]{F2F7FC}38.32 \\
\rowcolor[HTML]{FFFFFF} 
\cellcolor[HTML]{FFFFFF}Liu \& Heer & Y & 71.56 & \cellcolor[HTML]{F2F7FC}21.89 & 1.38 & \cellcolor[HTML]{F2F7FC}0.65 & 0 & \cellcolor[HTML]{F2F7FC}0 & 0.81 & \cellcolor[HTML]{F2F7FC}0.55 & 0 & \cellcolor[HTML]{F2F7FC}0 & 286.56 & \cellcolor[HTML]{F2F7FC}270.48 & 0 & \cellcolor[HTML]{F2F7FC}0 & 0 & \cellcolor[HTML]{F2F7FC}0 & 125.19 & \cellcolor[HTML]{F2F7FC}32.81 \\
\rowcolor[HTML]{EFEFEF} 
\cellcolor[HTML]{EFEFEF} & \cellcolor[HTML]{EFEFEF}A & 0 & \cellcolor[HTML]{E4EEFC}0 & 0 & \cellcolor[HTML]{E4EEFC}0 & 0 & \cellcolor[HTML]{E4EEFC}0 & 0 & \cellcolor[HTML]{E4EEFC}0 & 23.86 & \cellcolor[HTML]{E4EEFC}4.07 & 0 & \cellcolor[HTML]{E4EEFC}0 & 0 & \cellcolor[HTML]{E4EEFC}0 & 7.5 & \cellcolor[HTML]{E4EEFC}1.9 & 0 & \cellcolor[HTML]{E4EEFC}0 \\
\rowcolor[HTML]{EFEFEF} 
\cellcolor[HTML]{EFEFEF} & \cellcolor[HTML]{EFEFEF}BM & 22.58 & \cellcolor[HTML]{E4EEFC}3.27 & 0.79 & \cellcolor[HTML]{E4EEFC}0.26 & 0 & \cellcolor[HTML]{E4EEFC}0 & 5.79 & \cellcolor[HTML]{E4EEFC}1.72 & 21.42 & \cellcolor[HTML]{E4EEFC}3.09 & 22.21 & \cellcolor[HTML]{E4EEFC}4.28 & 0 & \cellcolor[HTML]{E4EEFC}0 & 10.04 & \cellcolor[HTML]{E4EEFC}3.07 & 19.29 & \cellcolor[HTML]{E4EEFC}3.36 \\
\rowcolor[HTML]{EFEFEF} 
\cellcolor[HTML]{EFEFEF} & \cellcolor[HTML]{EFEFEF}GZ & 8.5 & \cellcolor[HTML]{E4EEFC}2.86 & 3.71 & \cellcolor[HTML]{E4EEFC}1.49 & 0 & \cellcolor[HTML]{E4EEFC}0 & 3.33 & \cellcolor[HTML]{E4EEFC}1.48 & 6.54 & \cellcolor[HTML]{E4EEFC}2.4 & 20.08 & \cellcolor[HTML]{E4EEFC}3.85 & 0 & \cellcolor[HTML]{E4EEFC}0 & 9.29 & \cellcolor[HTML]{E4EEFC}2.8 & 18.88 & \cellcolor[HTML]{E4EEFC}3.28 \\
\rowcolor[HTML]{EFEFEF} 
\cellcolor[HTML]{EFEFEF}Wall & \cellcolor[HTML]{EFEFEF}Y & 1.83 & \cellcolor[HTML]{E4EEFC}0.55 & 1.42 & \cellcolor[HTML]{E4EEFC}0.28 & 0.17 & \cellcolor[HTML]{E4EEFC}0.15 & 1.12 & \cellcolor[HTML]{E4EEFC}0.21 & 0.12 & \cellcolor[HTML]{E4EEFC}0.24 & 22.88 & \cellcolor[HTML]{E4EEFC}4.13 & 7.04 & \cellcolor[HTML]{E4EEFC}1.92 & 0 & \cellcolor[HTML]{E4EEFC}0 & 18.75 & \cellcolor[HTML]{E4EEFC}3.28 \\
\bottomrule
\end{tabular}
\vspace{-2em}
\end{adjustbox}
\label{tab:seqtotals}
\end{table*}

\subsection{Analyzing Non-Terminal Taxonomies}
Similar to the analysis performed for terminal symbols, the coverage-based analysis of non-terminal symbols is analyzed by calculating the number of occurrences of established sequences or non-terminals in the interaction log datasets. And, the diversity-based analysis of non-terminal symbols is informed by finding new sequences or new non-terminals that are observed within and across interaction log datasets.

\begin{figure}[b]
    \centering
    \vspace{-8mm}
    \includegraphics[width=\columnwidth]{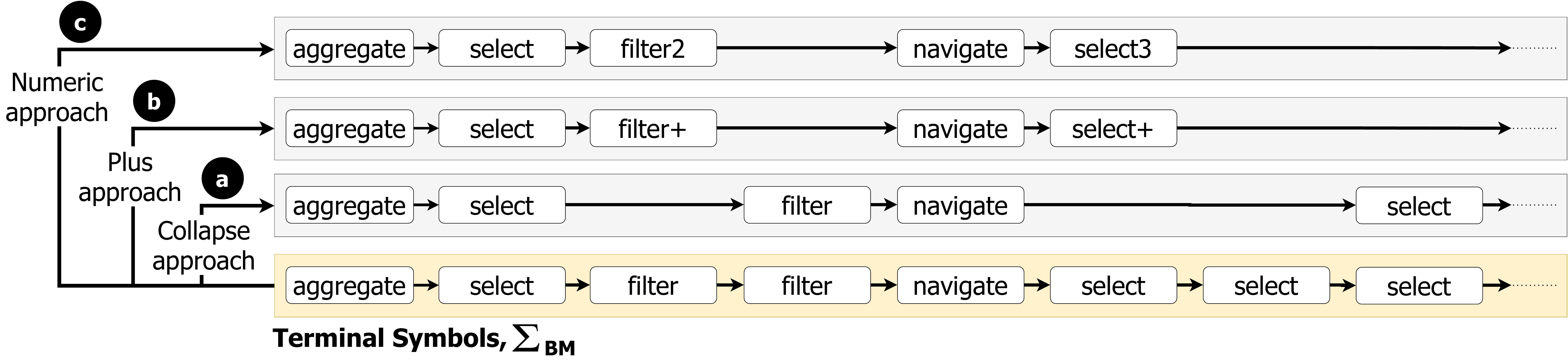}
    \caption{Different approaches to reduce the interaction log records of the interaction log datasets for data-driven exploration of new sequences: (a) Collapse, (b) Plus, (c) Numeric.}
    \label{fig:approach}
\end{figure}

\subsubsection{Coverage-based Analysis}
We calculate the coverage of non-terminal sequences observed in the interaction log datasets. The non-terminal symbols are sequences of interactions or patterns that recur within user analysis sessions, and can be represented using simpler terminal symbols. 
While calculating the number of non-terminals in the interaction log datasets, we observed multiple user sessions having consecutive log records that mapped to the same terminals. Since, having a sequence of similar terminals makes the chain of user terminals mappings very long and verbose for analysis, we use the ``collapse approach'' to collapse consecutive similar terminal symbols to one, as shown in \autoref{fig:approach} (a). The sequence of eight terminals are reduced to five terminals after applying the collapse approach on the \emph{filter} and \emph{select} repetitive terminals. We use the regular expressions generated for each of the non-terminal symbols to sum the number of sequences observed for each of the user sessions in the interaction log datasets. We report these in \autoref{tab:seqtotals} where we represent the average number of non-terminal sequences observed per user session with a confidence interval of $\pm$95\%. For instance, we observe an average of 2.25$\pm$0.55 $=$ 1.7 to 2.8, which is approximately 2 to 3 number of \emph{elaborating} Guo et al.~\cite{guo2015case} non-terminal sequences that are expressed using the Gotz \& Zhou (GZ)~\cite{gotz2009characterizing} underlying terminals in one user session of the Battle \& Heer~\cite{battle2019characterizing} interaction log dataset.

\begin{table*}[t]
\centering
\caption{New sequences observed using the \emph{plus} and \emph{numeric} approaches to analyzing the selected interaction log datasets.
}
\begin{adjustbox}{width=\textwidth,center}
\begin{tabular}{l|l|l|l|l|l}
\toprule
\multicolumn{1}{c|}{\textbf{\begin{tabular}[c]{@{}c@{}}Interaction Log Data\\\ Terminal Symbol\end{tabular}}} & \textbf{Approach} & \textbf{Amar et al.} & \textbf{Brehmer \& Munzner} & \textbf{Gotz \& Zhou} & \textbf{Yi et al.} \\
\midrule
\rowcolor[HTML]{EFEFEF} 
\cellcolor[HTML]{EFEFEF} & Plus & - & - & - & - \\
\rowcolor[HTML]{EFEFEF} 
\multirow{-2}{*}{\cellcolor[HTML]{EFEFEF}\textbf{Battle \& Heer}} & Numeric & - & - & - & - \\
\rowcolor[HTML]{FFFFFF} 
\cellcolor[HTML]{FFFFFF} & Plus &  & ((brush+, delete)+, brush+) & ((filter, navigate+)+, filter, navigate) & ((filter, explore+)+, filter, explore) \\
\rowcolor[HTML]{FFFFFF} 
\multirow{-2}{*}{\cellcolor[HTML]{FFFFFF}\textbf{Liu \& Heer}} & Numeric &  & (delete, brush) & ((navigate, change2)19, navigate, change) & ((connect, explore2)24, connect, explore) \\
\rowcolor[HTML]{EFEFEF} 
\cellcolor[HTML]{EFEFEF} & Plus & ((retrieve-value+, filter)+, retrieve-value) & (inspect+, annotate, inspect+) & (navigate+, select, navigate) & (explore+, select, explore+) \\
\rowcolor[HTML]{EFEFEF} 
\multirow{-2}{*}{\cellcolor[HTML]{EFEFEF}\textbf{Wall}} & Numeric & (filter, retrieve-value) & (change-metaphor) & (select, navigate) & (select, explore) \\
\bottomrule
\end{tabular}
\end{adjustbox}
\label{tab:diversitysequence}
\vspace{-4mm}
\end{table*}

We observe that the coverage of non-terminal symbols is less than that of the terminal symbols for our selected taxonomies and datasets. Of the representative non-terminal taxonomies, we observe the least coverage for the Guo et al.~\cite{guo2015case} non-terminals (maximum low counts of sequences), followed by Shneiderman's~\cite{shneiderman1996eyes} information-seeking mantra as we rarely observe any of the sequences in user sessions. On the other hand, the Gotz \& Wen~\cite{gotz2009behavior} non-terminal sequences seem to be more widely observed across user sessions. We attribute this behavior to the environments and tools used to come up with these non-terminals. For instance, since Guo et al.~\cite{guo2015case} realized the common user sequences only using a single and specific text analysis tool, we tend to not observe them in any other datasets. The Gotz \& Wen non-terminals seem to be the most prevalent across the interaction log datasets, making them more generalizable for multi-system and multi-task purposes.

\subsubsection{Diversity-based Analysis}
The analysis uses our regular grammars approach to find new common behaviors of users within and across the interaction log datasets. We use terminal mappings of individual taxonomies and split them per user sessions. Then we find intersection of consecutive sequence of terminals to attain the common sequences within the user sessions of an interaction log dataset. Similar to the previous analysis where we used the ``collapse'' approach to reduce the vast amounts of log data collected, we use the ``plus'' and ``numeric'' approach for this analysis, as shown in \autoref{fig:approach}. In the plus approach, consecutive similar terminal symbols are collapsed to one terminal symbol concatenated with a ``plus'' notation. For example, the consecutive \emph{filter} and \emph{select} terminals are replaced with \emph{filter+} and \emph{select+} terminals respectively, as seen in \autoref{fig:approach}. The numeric approach is similar to the plus approach but the plus notation is instead replaced with the number of times the terminal is repeated consecutively. Again as observed in \autoref{fig:approach}, the two consecutive \emph{filter} terminals and three consecutive \emph{select} terminals are replaced with \emph{filter2} and \emph{select3} respectively. We favor the plus and numeric approaches instead of the collapse approach for this analysis since the former approaches preserve the interactions of the users.

The new non-terminal sequences that we observed are shown in \autoref{tab:diversitysequence}. For the Battle \& Heer dataset, we do not observe any common terminal-level sequences across user sessions. For both the Liu \& Heer and Wall datasets, we observe different common non-terminal sequences for different terminal symbols as seen in \autoref{tab:diversitysequence}. Most of these sequences follow the same pattern of alternating between terminals and non-terminals. For instance, when using Amar et al. terminals, we observe that the Wall interaction log dataset alternates between multiple \emph{retrieve-value} and a \emph{filter} terminals followed by a \emph{retrieve-value} terminal. A similar pattern occurs with multiple \emph{brush} and a \emph{delete} terminals followed by a \emph{brush} terminal in the Liu \& Heer dataset (with Brehmer \& Munzner terminals). However, we see few exact pattern matches for the plus and numeric approaches, suggesting that even when users perform similar patterns, the number of interactions (or terminals) within these patterns often varies. We further extended this analysis to search for common sequences \emph{across} all three datasets, but we did not find any universal patterns. Thus, the studied non-terminals appear to be less expressive.
\section{Discussion}
\label{sec:discussion}

The terminal-level taxonomies we studied have mixed \emph{expressiveness} as defined by the \emph{coverage} and \emph{diversity} of the mapped terminals. We observe high coverage for these taxonomies across all three log datasets, suggesting that the current set of low-level taxonomies in the literature provide sufficient coverage of log records within real-world interaction log datasets. However, we find that some of the representative terminal taxonomies tend to under- or over-utilize certain terminals, resulting in skewed distributions of emitted terminals within our mappings and thus limited diversity. 

We believe the mixed results stem from the tension between optimizing for \emph{specificity}, or ensuring specific user activities are represented, and \emph{generality}, or designing terminals that can be applied to many tools, when designing taxonomies. On the one hand, our findings demonstrate the utility of popular theoretical taxonomies beyond their intended use as descriptive tools for designers. Our grammar-based approach reveals how taxonomies can be useful tools for analyzing interaction logs from a variety of systems. On the other hand, our approach demonstrates the limitations of taxonomies. For example, our results suggest that highly skewed taxonomies may be too general, sacrifice context, and as a result lead to mappings that may not be meaningful for interaction log analysis. In some cases, these taxonomies produced a single terminal for most (e.g., over 95\% of) log records, resulting in homogeneous mappings. However, striving purely for specificity may also produce taxonomies with limited applicability, e.g., taxonomies that apply only to one tool. Our research highlights a potential need for more taxonomies that strike a balance between specificity and generality. It also suggests a potential direction for achieving this balance: augmenting taxonomies to include critical contextual cues as input to the underlying regular grammar; for example, details about the system wherein the interactions are being performed.

\begin{figure}[b]
    \centering
    \vspace{-6mm}
    \includegraphics[width=\columnwidth]{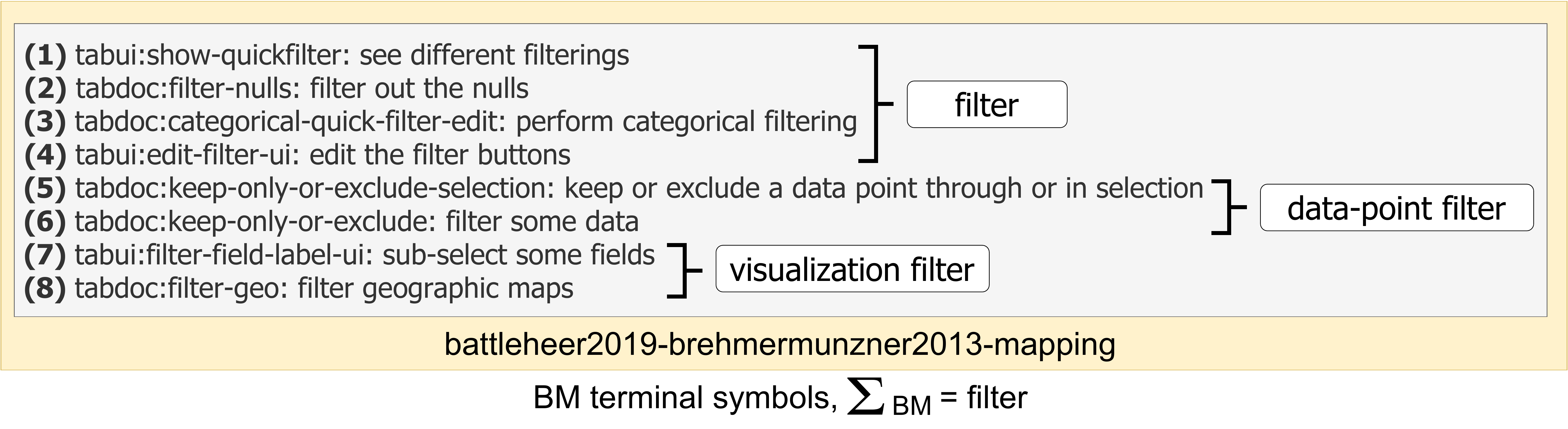}
    \caption{Eight Tableau log records mapped to the same terminal in the Brehmer \& Munzner taxonomy (\emph{filter}), which could be represented as three terminals: \emph{filter}, \emph{data-point filter}, and \emph{visualization filter}.}
    \label{fig:discussion}
\end{figure}

In contrast, we rarely observe the interaction patterns proposed in our selected non-terminal taxonomies, and the few we do observe are only a small fraction of the analyzed logs. Although these taxonomies do not directly match the interaction log datasets we analyzed, our regular grammars approach enables deducing new \emph{data-driven taxonomies} at higher levels of user activity such as interaction sequences and analysis tasks. For example, our approach reveals common sub-sequences within interaction logs which represent more meaningful and complex patterns than those proposed in well-known taxonomies. Finally, we do not observe any common sequences that occur across all three log datasets. We believe these issues stem in part from a mismatch between popular non-terminal taxonomies and log recording strategies, as well as challenges originating from the terminal rather than non-terminal level: over- and under-utilization of certain terminals and lack of important contextual cues in these taxonomies. 

However, only a subset of taxonomies and log datasets are analyzed in this work. We encourage the community to extend these ideas to see how they generalize to other taxonomies and log analysis contexts.

\textbf{Implications for Log Data Analysis.} There are two important concerns when applying taxonomies to interaction logs. First is the loss of information in the resulting mappings such as a lack of proper translation between the context or \emph{semantics} of the user action and taxonomy category or \emph{syntax}. Taxonomies are designed to concisely communicate the semantics of user interactions. However, a user's interaction intent is also influenced by the design of the underlying interface, which is intentionally abstracted away from most taxonomies. As an example, consider the \texttt{filter} terminal from the Brehmer \& Munzner taxonomy. While mapping the Battle \& Heer interaction log dataset, we notice that \emph{eight} of its distinct log records were mapped to the \texttt{filter} terminal, as shown in \autoref{fig:discussion}. After investigating the underlying system details, we posit that these eight filters actually represent \emph{three} types of filtering interactions, shown in in \autoref{fig:discussion}: filter the data, filter the visualization, and ``other'' filter operations. These findings point to a need to \emph{augment taxonomies} to include system level details to prevent losing user context. A second concern lies in the inability to express timing of interactions because of the use of grammar-based approach. However, to facilitate the development of new taxonomies, we need to ensure that log data collection processes \emph{scale}, which is generally enabled by developing structured languages, as observed in other areas such as databases, distributed systems, and NLP.

Further, although many interaction logs are shared online, this is no guarantee that others will actually be able to use them. It is critical to create a community-wide process for sharing datasets that will be reusable~\cite{battle2018evaluating}. Logs are often collected in an ad-hoc manner that is unique to the system being evaluated, making it difficult to translate these logs to a broad range of analysis contexts~\cite{psallidas2018smoke,cutler2020trrack}. Generalizable logging formats must be adopted to make future log datasets applicable to a wider range of analysis scenarios. We need a shift in how we think about data sharing and consider data useful only if others can and actually \emph{use} it.

\section{Conclusion}
This paper presents a framework that bridges the gap between theoretical visualization task taxonomies and empirical analysis of interaction log data. to do this, we exploit structural parallels between how people express themselves through interactions and language by reformulating existing theories as regular grammars. We represent interactions as terminals within a regular grammar and patterns of interactions as regular expressions over these terminals to capture common language patterns. Regular grammars provide opportunities to express new taxonomies in exciting ways. For example, this formulation can enable future work to express new taxonomies as a mix of low-level and high-level attributes of the inherently hierarchical structure of visual analysis and human reasoning. Our contributions can help the community to create taxonomies that match a broader range of granularities in user intents and try to strike a subtler balance between capturing the coverage and diversity of interaction log events.

\section{Acknowledgements}
This project was partially supported by the National Science Foundation under Grants OAC-2118201 and  IIS-1850115.

\clearpage
\printbibliography                
\end{document}